\documentclass[showpacs,preprintnumbers,prd,nofootinbib,floats,amssymb,floatfix]{revtex4}
\usepackage{amsmath}
\usepackage{amsfonts}
\usepackage{graphicx}
\usepackage{hyperref}

\usepackage{amsmath}
\usepackage{amstext}
 
\begin{document}

\title { A pure Dirac's method for  Yang-Mills  expressed as  a constrained BF-like  theory}  
\author{ Alberto Escalante}  \email{aescalan@sirio.ifuap.buap.mx}
 \affiliation{Instituto de F{\'i}sica Luis Rivera Terrazas, Benem\'erita Universidad Aut\'onoma de Puebla, (IFUAP).
   Apartado postal      J-48 72570 Puebla. Pue., M\'exico, } 
   \author{J. Berra}
 \affiliation{ Facultad de Ciencias F\'{\i}sico Matem\'{a}ticas, Benem\'erita Universidad Au\-t\'o\-no\-ma de Puebla,
 Apartado postal 1152, 72001 Puebla, Pue., M\'exico.}

\begin{abstract}
A {\it pure Dirac's method} of Yang-Mills  expressed as a constrained $BF$-like theory is performed.  
In this paper we study  an action principle composed by the coupling of two topological  $BF$-like theories,  which at the Lagrangian level reproduces  Yang-Mills  equations.  By a pure Dirac's method we mean that we consider all the variables that occur in the Lagrangian density as dynamical variables and not only those ones that involve temporal derivatives. The analysis in the complete phase space enable us to calculate
the extended Hamiltonian, the extended action, the constraint algebra, the gauge transformations and then we carry out the counting of degrees of freedom. We show that the  constrained $BF$-like theory correspond at classical level to Yang-Mills theory. From the results obtained,  we discuss briefly the quantization of the theory. In addition we compare our results with alternatives models that have been reported in the literature.\\  

\end{abstract}
\date{\today}
\preprint{}
\maketitle


\section{ INTRODUCTION}

Nowadays,  the study of topological field theories is a topic of great interest in physics. The importance for studying  those
 theories lies in a closed relation with physical theories as  for instance,  Yang-Mills  [YM] and  General Relativity \cite{carlo, tomas}. Topological field  theories  are  characterized
 by being devoid of local physical degrees of freedom\footnote{In the paper we refer as a topological theory,  a classical theory lacking of  local degrees of freedom, even though it does possess global physical degrees of freedom, which are characteristic by means of  the topological properties either of the internal field space or of the base spacetime manifold.} ,   they are background independent  and  diffeomorphisms covariant  
 \cite{3a, 4a}.  Relevant  examples of topological field theories are  the  
 so called $BF$ theories. $BF$  theories  were introduced as generalizations of three dimensional Chern-Simons actions and in a certain sense this is 
the simplest possible gauge theory. It can be defined on spacetimes of any dimension. It is background free, meaning that to formulate it we do not need
a pre-existing metric or any other such geometrical structure on spacetime \cite{5a, 6a}. At classical level, the theory has no local degrees of freedom; all the 
interesting observables are global in nature  and this seems to remain true upon quantization \cite{baez}.  Thus $BF$ theory serves as a simple starting point for the 
studies of background free theories. In particular, general relativity in 3-dimensions is a special case of $BF$ theory, while general relativity in 4-dimensions can be viewed as a $BF$ theory with extra constraints \cite{8aa}. Furthermore,  we are able to find in the
 literature several examples where $BF$ theories come to be relevant models  for instance,  in  alternative   
 formulations of gravity such as the MacDowell-Mansouri approach \cite{7a}.  MacDowell-Mansouri formulation of gravity consists 
in  breaking down  the internal symmetry group of a  $BF$-theory from $SO$(5) to $SO$(4),  to obtain   Palatini's 
action plus a sum of the second Chern  and Euler topological invariants. Because these terms have trivial 
local variations  that do not contribute classically to the dynamics, one thus obtain
 essentially general relativity. On the other hand,  within the framework of [YM] theories, we can find some 
cases where  $BF$ theories have been relevant, an  example of this is  Martellini's model \cite{8a}. This model 
consists in to express [YM] theory as a $BF$-like theory in order to expose its relation with topological $BF$ theory.
 Thus, the first-order formulation (BF-YM) is equivalent 
on shell to the usual second-order formulation (YM). In fact, after a Wick rotation both formulations of the theory possess the same 
perturbative quantum properties; the Feynman rules, the structure of one loop divergent diagrams  and renormalization
 has been studied founding that there exists an equivalence of the $uv$-behaviour for both approaches \cite{Martellini 1}. The main advantage of this formulation
lies in the possibility to express some observables like the color magnetic operator in the continuum and express it as the dual 't Hooft observable
employing the abelian projection gauge \cite{Cattaneo}. Nevertheless, the canonical quantization perspective is more subtle under Wick rotations, because
it shows that in order to make the theory euclidean, we must to complexify the Poisson algebra \cite{Martellini 2}. Recently other approaches use a well known duality between Maxwell-Chern Simons theory and a self dual massive model, this description has been extended to topological massive gauge theories providing a topological mechanism to generate mass for the bosonic $p$-tensor fields in any spacetime dimension \cite{Bertrand,Bertrand1}.\\
On the other side,  we are able to find that  Lisi in \cite{Lisi} and Smolin in \cite{Smolin} have  worked  with  $BF$-like theories written  as an extension of    Plebanski's action, and in those works  they  got a consistent dynamics for any group $G$ containing the local Lorentz group, and then by using a simple mechanism which breaks down  the symmetry, they   obtain as resulting   dynamics  to  [YM] coupled to general relativity plus corrections. \\
At the light of these facts, in this paper we analyze a  $BF$-like action yielding  [YM] equations of motion. We study  the principal symmetries of that action by means of a detailed canonical analysis    using a pure Dirac's method,  and  the quantization procedure is discussed. With the terminology \textit{  a pure Dirac's method} we mean that we shall   consider   in the Hamiltonian framework  all the fields that define our theory are dynamical ones. Of course, our approach differs from the standard Dirac's analysis,  because  the standard analysis is  developed on a smaller phase space by  considering as dynamical variables only those variables  with time derivative occurring explicitly in the Lagrangian. In this paper,  our approach present clear advantages in respect to the standard one,  namely;  by working on the full phase space, we will able to know the full structure of the constrains and their algebra, the equations of motion obtained from the extended action  and the full structure of  the gauge transformations as well.  The approach used in the present paper, has been performed    to  diffeomorphism covariant   field theories \cite{4aa, 4bb},  showing results that are not obtained by means of a standard Dirac's analysis. In those works    were   reported the full structure of the constraints on the full phase space for the Second-Chern class and  the latter for general relativity  in the $G \rightarrow 0$ limit, being $G$  the gravitational coupling constant. The correct identification of the constraints is a very important step because are used to carry out the counting of the physical degrees of freedom and they let us to know  the gauge transformations if there exist first class constraints. On the other hand, the constraints are the guideline to make the best progress for the quantization of the theory, therefore it  is mandatory to know their full structure \cite{4a}. It is worthwhile to mention,  that  the constraints obtained  by performing  a pure Dirac's formalism,  the algebra among them is closed,  and is not necessary to fix by hand the constraints as in the case of Plebanski theory \cite{Peldan, ww},  because the method itself provides us the required structure. One example of  ambiguities  found by developing  the hamiltonian analysis on a reduced phase space,  is presented  in  three dimensional tetrad gravity, in despite of the existence of several  articles performing the hamiltonian analysis, in some papers
it is written that the gauge symmetry is Poincare symmetry \cite{Witten}, in others that is Lorentz symmetry plus diffeomorphisms \cite{Carlip}, or that  there exist various ways to define the constraints leading to different gauge transformations. We think that the complete Hamiltonian method ( a pure Dirac's formalism) is the best  tool  for solving those problems.  \\
Finally we show that the action analyzed in this paper is  the coupling of topological theories namely;  the  $BF$-like action studied here can be split  in  two terms  lacking   of physical  degrees of freedom, the complete action,  however, does has physical degrees of freedom, the [YM] degrees of freedom. \\ 
The paper is organized as follows: In section II,  we show that  [YM]  action differ from   the $BF$-like action studied here, because it  is expressed as the coupling  of two topological terms just as General Relativity in Plebanski's formulation \cite{ww}.  By a pure Dirac's method of the $BF$-like theory, one realize  that the couplet terms incorporate  reducibility conditions in the  constraints, thus, the topological invariance of the full action will be  broken  emerging degrees of freedom. In Section III, a complete canonical analysis  analysis of   Martellini's model is performed, this exercise has  not  reported in the literature, then we compare the results of this section  with those obtained in  previous sections.  We finish with  some remarks about our results. 
\newline
\setcounter{equation}{0} \label{c2}
\section{A pure Dirac's method for   [YM] theory  expressed  as a constrained  BF-like  theory}
The action of our interest is given by
\begin{equation}
S[A,B]= \int_M \ast  B^{I} \wedge B^{I} -2 B^{I}\wedge \ast F^{I},
\label{bf-ym1}
\end{equation}
where $F^{I}$ corresponds to the curvature of the connection
one-form $A^{I}$ valued on the algebra of $SU(N)$. In this manner, the action
(\ref{bf-ym1}) takes the form
\begin{equation}
S[B,A]=\int_{M}\frac{1}{4}B_{\mu\nu}^{I}B^{\mu\nu
I}-\frac{1}{2}B^{\mu\nu
I}\left(\partial_{\mu}A^I_{\nu}-\partial_{\nu}A_{\mu}^{I}+f^{IJK}A_{\mu}A_{\nu}\right),
\label{bf-ym}
\end{equation}
\noindent the equations of motion obtained from (\ref{bf-ym}) are given by
\begin{equation}
B_{\mu\nu}^{I}=F_{\mu\nu}^{I},\; \;\;\;\;\;\;\; D_{\mu}B^{\mu\nu
I}=0,
\end{equation}
\noindent which correspond  to [YM] equations of motion. By substituting
the former equations of motion in the action we recover the [YM]
action
\begin{equation}
S[A]=-\int_{M}\frac{1}{4}F_{\alpha\beta}^{I}F^{\alpha\beta
}_ Id^{4}x,
\label{eq.4}
\end{equation}
\noindent where
$F_{\alpha\beta}^{I}=\partial_{\mu}A^I_{\nu}-\partial_{\nu}A_{\mu}^{I}+f^{I}{_{JK}}A^J_{\mu}A^K_{\nu}$,
is  the curvature tensor valued on a Lie algebra. It is important to remark, that the role of the dynamical variables occurring  in  the actions (\ref{bf-ym}) and (\ref{eq.4})  is  quite different. For the former, $A_\mu$ and $B_{\mu \alpha}$  both are   determined by the dynamics. For the later, $A_\mu$  is determined by the dynamics and $F_{\mu \alpha}$ is not a dynamical variable anymore, it is a label.  \\
In order to procedure with our analysis, we are able to  observe  that if we split the
action (\ref{bf-ym}) in two parts, say
\begin{equation}\label{S1}
S_{1}[B]=\int_{M}\frac{1}{4}B_{\mu\nu}^{I}B^{\mu\nu }_I,
\end{equation}
\noindent and
\begin{equation}\label{S2}
S_{2}[A,B]=\int_{M}\frac{1}{2}B_ I^{\mu\nu
}\left(\partial_{\mu}A_{\nu}^{I}-\partial_{\nu}A_{\mu}^{I}+f^I_{{JK}}A_{\mu}^{J}A_{\nu}^{K}\right),
\end{equation}
\noindent we obtain two topological field theories. In fact, we can
see immediately that $S_{1}[B]$ is topological since does not have
dynamical variables occurring  in the action. On the other hand, we will show below that the action
(\ref{S2}) is a topological one as well. May be for the lector this part is not relevant, however, we need to remember that topological field theories are characterized by being devoid of
local degrees of freedom. That is, the theories are susceptible only
to global degrees of freedom associated with non-trivial topologies
of the manifold in which they are defined and topologies of the
gauge bundle \cite{21a, 22a}. Thus,   it is mandatory to perform the canonical analysis of the action 
(\ref{S2}) because there is a gauge group.     \\
 So, 
 by performing the 3+1 decomposition in (\ref{S2}) we
obtain
\begin{equation}
S_{2}[A,B]=\int\int_{\Sigma}\left[B^{0iI}(\dot{A}_{i}^{I}-\partial_{i}A_{0}^{I}+f^{IJK}A_{0}^{J}A_{i}^{K})+\frac{1}{2}B^{ijI}F_{ij}^{I}\right],
\label{49n}
\end{equation}
\noindent where
$F_{ij}^{I}=\partial_{i}A_{j}^{I}-\partial_{j}A_{i}^{I}+f^{I}{_{JK}}A_{i}^{J}A_{j}^{K}$.  \\
\noindent Dirac's method calls for the definition of the momenta
$(\Pi^{\alpha I},\Pi^{\alpha\beta I})$ canonically conjugate to the
dynamical variables $(A_{\alpha}^{I},B_{\alpha\beta}^{I})$
\begin{equation}
\Pi^{\alpha I}=\frac{\delta \mathcal{L}}{\delta
\dot{A}_{\alpha}^{I}}, \;\;\;\;\; \Pi^{\alpha\beta
I}=\frac{\delta\mathcal{L}}{\delta \dot{B}_{\alpha\beta}^{I}},
\end{equation}
\noindent On the other hand, the matrix elements of the Hessian
\begin{equation}
\frac{\partial^{2}\mathcal{L}}{\partial(\partial_{\mu}A_{\alpha}^{I})\partial(\partial_{\mu}A_{\rho}^{J})},\;\;\;
\frac{\partial^{2}\mathcal{L}}{\partial(\partial_{\mu}B_{\alpha\beta}^{I})\partial(\partial_{\mu}A_{\rho}^{J})},\;\;\;
\frac{\partial^{2}\mathcal{L}}{\partial(\partial_{\mu}B_{\alpha\beta}^{I})\partial(\partial_{\mu}B_{\rho\gamma}^{J})},
\end{equation}
\noindent are identically zero, thus the rank of the Hessian is zero.
Therefore,  we expect $10(N^{2}-1)$ primary constraints. From the
definition of the momenta we identify the following  10 primary
constraints
\begin{equation}\begin{split}
\phi^{0I}:\Pi^{0I}&\approx 0,\\
\phi^{iI}:\Pi^{iI}-B^{0iI}&\approx 0,\\
\phi^{0i}_I:\Pi^{0i}_I&\approx 0,\\
\phi^{ij}_I:\Pi^{ij}_I&\approx 0,
\end{split}
\end{equation}
\noindent The canonical Hamiltonian density for the system has the
following form
\begin{equation}\begin{split}
\mathcal{H}_{c}&=\dot{A}_{\mu}^{I}\Pi^{\mu
}_I+\dot{B}_{0i}^{I}\Pi^{0i}_I+\dot{B}_{ij}^{I}\Pi^{ij}_I-\mathcal{L}\\
&=-A_{0}^{I}D_{i}\Pi^{0i}_I-\frac{1}{2}B^{ij}_IF_{ij}^{I},
\end{split}
\end{equation}
\noindent
Thus,  by taking in to account the primary constraints, we can identify
the primary Hamiltonian given by
\begin{equation}
H_{P}=H_{c}+\int
d^{3}x\left[\lambda_{0}^{I}\phi^{0}_I+\lambda_{i}^{I}\phi^{i}_I+\lambda_{0i}^{I}\phi^{0i}_I+\lambda_{ij}^{I}\phi^{ij}_I\right],
\end{equation}
\noindent where
$\lambda_{0}^{I},\lambda_{i}^{I},\lambda_{0i}^{I},\lambda_{ij}^{I}$
are Lagrange multipliers enforcing the constraints. The fundamental
Poisson brackets for our theory are given by
\begin{equation}\begin{split}
\{A_{\alpha}^{I}(x),\Pi^{\mu
I}(y)\}&=\delta^{\mu}_{\alpha}\delta^{IJ}\delta^{3}(x-y),\\
\{B_{\alpha\beta}^{I}(x),\Pi^{\mu\nu
I}(y)\}&=\frac{1}{2}\left(\delta_{\alpha}^{\mu}\delta_{\beta}^{\nu}-\delta_{\beta}^{\mu}\delta_{\alpha}^{\nu}\right)\delta^{3}(x-y).
\end{split}
\end{equation}
\noindent In this manner, by using the fundamental Poisson brackets
for our theory we find that the following $10(N^{2}-1)\times
10(N^{2}-1)$ matrix whose entries are the Poisson brackets among
the primary constraints
\begin{eqnarray}
\{ \phi^{0}_I(x),\phi^{0}_J(y) \}&=&0,   \qquad   \{ \phi^{0}_I(x),\phi^{i}_J(y) \} = 0,  \nonumber \\
\{ \phi^{0}_I(x),\phi^{0i}_J(y) \}&=&0,   \qquad   \{ \phi^{0}_I(x),\phi^{ij}_J(y) \} = 0,  \nonumber \\
\{ \phi^{i}_I(x),\phi^{j}_J(y) \}&=&0,   \qquad   \{ \phi^{i}_I(x),\phi^{0j}_J(y) \} = \frac{1}{2}\delta{i}_{j}\delta_{IJ}\delta^{3}(x-y),  \nonumber \\
\{ \phi^{i}_I(x),\phi^{jk}_J(y) \}&=&0,   \qquad   \{ \phi^{ij}_I(x),\phi^{kl}_J(y) \} = 0,  \nonumber 
\end{eqnarray}
\noindent has rank $6(N^{2}-1)$ and $(4(N^{2}-1))$ null vectors.
This means that we expect $4(N^{2}-1)$ secondary constraints
\begin{equation}\begin{split}
\dot{\phi^{0}_I}&=\{\phi^{0}_I,H_{P}\}\approx 0 \;\; \Rightarrow \;\;
\psi_{I}:=D_{i}\Pi^{i}_I\approx 0,\\
\dot{\phi^{ij}_I}&=\{\phi^{ij}_I,H_{P}\}\approx 0 \;\; \Rightarrow
\;\; \psi_{ij}^I:=\frac{1}{2}F_{ij}^I\approx 0,
\end{split}
\end{equation}
\noindent and the rank allows us to fix the following Lagrange
multipliers
\begin{equation}\begin{split}
\dot{\phi^{0iI}}&=\{\phi^{0iI},H_{P}\}\approx 0 \;\; \Rightarrow\;\;
\lambda_{i}^{I}=0,\\
\dot{\phi^{ijI}}&=\{\phi^{ijI},H_{P}\}\approx 0 \;\; \Rightarrow\;\;
\lambda_{0i}^{I}=2D_{j}B^{ij}_I+2f{_{I}}^{JK}A_{0}^{J}\Pi^{iK}.
\end{split}
\end{equation}

\noindent For this theory there are not, third constraints. In this
manner, with all the constraints at hand, we need to identify those
that are first and second class kind. For this purpose, we can
observe that the $14(N^{2}-1)\times 14(N^{2}-1)$ matrix whose
entries are the Poisson brackets among the primary and secondary
constraints given by

\begin{eqnarray}
\{ \phi^{0P}(x),\phi^{0I}(y) \}&=&0,   \qquad   \{ \phi^{0P}(x),\phi^{iI}(y) \} = 0,  \nonumber \\
\{ \phi^{0P}(x),\phi^{0iI}(y) \}&=&0,   \qquad   \{ \phi^{0P}(x),\phi^{ijI}(y) \} = 0,  \nonumber \\
\{ \phi^{lP}(x),\phi^{iI}(y) \}&=&0,   \qquad   \{ \phi^{lP}(x),\phi^{0iI}(y) \} = \frac{1}{2}\delta^{i}_{l}\delta^{PI}\delta^{3}(x-y),  \nonumber \\
\{ \phi^{lP}(x),\phi^{ijI}(y) \}&=&0,   \qquad   \{ \phi^{lmP}(x),\phi^{ijI}(y) \} = 0,  \nonumber \\
\{ \phi^{0P}(x),\psi^{I}(y) \}&=&0,   \qquad   \{ \phi^{lP}(x),\psi^{I}(y) \} = f^{PIK}\Pi^{lK}\delta^{3}(x-y),  \nonumber \\
\{ \phi^{0lP}(x),\psi^{I}(y) \}&=&0,   \qquad   \{ \phi^{lmP}(x),\psi^{I}(y) \} = 0,  \nonumber \\
\{ \psi^{P}(x),\psi^{I}(y) \}&=&f^{PIK}D_{i}\Pi^{iK},   \qquad   \{ \psi^{lmP}(x),\psi^{I}(y) \} = 0,  \nonumber \\
\{ \phi^{0P}(x),\psi^{ijI}(y) \}&=&0,   \qquad   \{ \phi^{lP}(x),\psi^{ijI}(y) \} =\frac{1}{2}\left(-\delta^{l}_{j}\delta^{PI}\partial_{i}+\delta^{l}_{i}\delta^{PI}\partial_{j}+f^{PIK}(\delta^{l}_{i}A_{j}^{K}-\delta^{l}_{j}A_{i}^{K})\right)\delta^{3}(x-y),  \nonumber \\
\{ \phi^{lmP}(x),\psi^{ijI}(y) \}&=&0,   \qquad   \{ \psi^{lmP}(x),\psi^{ijI}(y) \} = 0,  \nonumber 
\end{eqnarray}
\noindent has rank=$6(N^{2}-1)$ and $8(N^{2}-1)$ null-vectors. From
the null-vectors it is possible to  identify the following $8(N^{2}-1)$ first class
constraints
\begin{equation}\begin{split}
\gamma^{0}_I&=\Pi^{0}_I\approx 0,\\
\gamma^{ij}_I&=\Pi^{ij}_I\approx 0,\\
 \gamma^{I}&=D_{i}\Pi^{iI}+2f^{I}{_{JK}}B_{0i}^{J}\Pi^{0iK},\\
 \gamma_{ij}^I&=\frac{1}{2}F_{ij}^{I}+\frac{1}{2}[D_{i}\Pi{^0}{_{j}}^{I}-D_{j}\Pi{^0}{_{i}}^{I}].
\end{split}
\label{65v}
\end{equation}
\noindent 
On the other side, the rank allows us to identify the next
$6(N^{2}-1)$ second class constraints
\begin{equation}\begin{split}
\chi^{0i}_I&=\Pi^{0i}_I\approx 0,\\
\chi^{i}_I&=\Pi^{i}_I-B^{0i}_I\approx 0.
\end{split}
\end{equation}

\noindent However, we can observe in (\ref{65v}) that the fourth constraint   can be written as
\begin{equation}
\Upsilon_{i}^I\equiv
\eta_{i}{^{jk}}\gamma_{jk}^{I}=\frac{1}{2}\eta_{i}{^{jk}}F^{I}_{jk}+2\eta_{i}{^{jk}}D_{j}\Pi{^0}{_{k}}^{I},
\end{equation}

\noindent thus, $D_{i}\Upsilon^{iI}-\eta_{k}{^{ij}}f^{I}{_{JK}}F_{ij}^{J}\Pi^{0kK} =0$ because of
Bianchi's identity $\eta^{ijk}D_{i}F^{I}_{jk}=0$. In this way, the former  relation
represents a reducibility condition. This means that the number of
independent first class constraints corresponds to $[8-1](N^2-1)=7(N^2-1)$. Therefore, the counting of degrees of freedom can be carry out as
follows: There are $20(N^{2}-1)$ canonical variables, $7(N^2-1)$ independent
first class constraints and $6(N^{2}-1)$ independent second class
constraints. Therefore, the system expressed by the action principle
(\ref{bf-ym}) is devoid of physical degrees of freedom and corresponds to be 
a topological theory. In this manner, the separated actions (\ref{S1}) and (\ref{S2})
are  topological field theories. \\
 Now we will perform a pure Dirac's analysis for the
coupled  action (\ref{bf-ym}) and we will show that will be not topological anymore. For our aims,  we perform the $3+1$ decomposition of (\ref{bf-ym}) obtaining 

\begin{equation}
S[A,B]=\int\int_{\Sigma}\left[\frac{1}{2}B_{0i}^{I}B^{0iI}+\frac{1}{4}B_{ij}^{I}B^{ijI}-B^{0iI}(\dot{A}_{i}^{I}-\partial_{i}A_{0}^{I}+f^{ijk}A_{0}^{J}A_{i}^{K})-\frac{1}{2}B^{ijI}F_{ij}^{I}\right]d^{3}xdt,
\end{equation}
 thus, to perform a pure Dirac's method we need the
definition of the momenta $(\Pi^{\alpha I},\Pi^{\alpha\beta I})$
canonically conjugate to $(A_{\alpha}^{I},B_{\alpha\beta}^{I})$
\begin{equation}\label{momenta bf}
\Pi^{\alpha I}=\frac{\delta \mathcal{L}}{\delta
\dot{A}_{\alpha}^{I}}, \;\;\;\;\; \Pi^{\alpha\beta I}=\frac{\delta\mathcal{L}}{\delta \dot{B}_{\alpha\beta}^{I}},
\end{equation}

\noindent on the other hand, the matrix elements of the Hessian
\begin{equation}
\frac{\partial^{2}\mathcal{L}}{\partial
\dot{A}_{\alpha}^{I}\partial\dot{A}_{\rho}^{J}},\;\;\;
\frac{\partial^{2}\mathcal{L}}{\partial\dot{B}_{\alpha\beta}^{I}\partial\dot{A}_{\rho}^{J}},\;\;\;
\frac{\partial^{2}\mathcal{L}}{\partial\dot{B}_{\alpha\beta}^{I}\partial\dot{B}_{\rho\gamma}^{J}},
\end{equation}
\noindent are identically zero, the rank of the Hessian is zero.
Thus, we expect $10(N^{2}-1)$ primary constraints. From the
definition of the momenta (\ref{momenta bf}), we identify the next $10(N^{2}-1)$
primary constraints
\begin{equation}\begin{split}
\phi^{0I}&:\Pi^{0I}\approx 0,\\
\phi^{iI}&:\Pi^{iI}+B^{0iI}\approx 0,\\
\phi^{0iI}&:\Pi^{0iI}\approx 0,\\
\phi^{ijI}&:\Pi^{ijI}\approx 0.
\end{split}
\end{equation}
\noindent The canonical Hamiltonian density for the system has the
next form
\begin{equation}\begin{split}
\mathcal{H}_{c}&=\dot{A}_{\mu}^{I}\Pi^{\mu
I}+\dot{B}_{\mu\nu}^{I}\Pi^{\mu\nu I}-\mathcal{L}\\
&=\frac{1}{2}\Pi^{iI}\Pi_{i}^{I}-\frac{1}{4}B_{ij}^{I}B^{ijI}-A_{0}^{I}D_{i}\Pi^{iI}+\frac{1}{2}B^{ijI}F_{ij}^{I}.
\end{split}
\end{equation}
\noindent Thus the primary Hamiltonian is given by
\begin{equation}
H_{P}=H_{c}+\int
d^{3}x\left[\lambda_{0}^{I}\phi^{0I}+\lambda_{i}^{I}\phi^{iI}+\lambda_{0i}^{I}\phi^{0iI}+\lambda_{ij}^{I}\phi^{ijI}\right],
\end{equation}
\noindent where
$\lambda_{0}^{I},\lambda_{i}^{I},\lambda_{0i}^{I},\lambda_{ij}^{I}$
are Lagrange multipliers enforcing the constraints. The fundamental
Poisson brackets for our theory are given by
\begin{equation}\begin{split}
\{A_{\alpha}^{I}(x),\Pi^{\mu
J}(y)\}&=\delta^{\mu}_{\alpha}\delta^{IJ}\delta^{3}(x-y),\\
\{B_{\alpha\beta}^{I}(x),\Pi^{\mu\nu
J}(y)\}&=\frac{1}{2}\left(\delta_{\alpha}^{\mu}\delta_{\beta}^{\nu}-\delta_{\beta}^{\mu}\delta_{\alpha}^{\nu}\right)\delta^{IJ}\delta^{3}(x-y).
\end{split}
\end{equation}
\noindent The $10(N^{2}-1)\times 10(N^{2}-1)$ matrix whose entries
are the Poisson brackets among the primary constraints are given by
\begin{eqnarray}
\{ \phi^{0P}(x),\phi^{0I}(y) \}&=&0,   \qquad   \{ \phi^{0P}(x),\phi^{iI}(y) \} = 0,  \nonumber \\
\{ \phi^{0P}(x),\phi^{0iI}(y) \}&=&0,   \qquad   \{ \phi^{0P}(x),\phi^{ijI}(y) \} = 0,  \nonumber \\
\{ \phi^{lP}(x),\phi^{iI}(y) \}&=&0,   \qquad   \{ \phi^{lP}(x),\phi^{0iI}(y) \} = -\frac{1}{2}\delta^{l}_{i}\delta^{PI}\delta^{3}(x-y),  \nonumber \\
\{ \phi^{lP}(x),\phi^{ijI}(y) \}&=&0,   \qquad   \{ \phi^{0lP}(x),\phi^{0iI}(y) \} = 0,  \nonumber \\
\{ \phi^{0lP}(x),\phi^{ijI}(y)\}&=&0,   \qquad
\{\phi^{lmP}(x),\phi^{ijI}(y) \}= 0, \nonumber 
\end{eqnarray}
\noindent has rank $6(N^{2}-1)$ and $4(N^{2}-1)$ null vectors.
Thus by using the null vectors, consistency conditions yield  the
following $4(N^{2}-1)$ secondary constraints
\begin{equation}\begin{split}
\dot{\phi}^{0I}&=\{\phi^{0I},H_{P}\}\approx 0 \;\; \Rightarrow \;\;
\psi^{I}:=D_{i}\Pi^{iI}\approx 0,\\
\dot{\phi}^{ijI}&=\{\phi^{ijI},H_{P}\}\approx 0 \;\; \Rightarrow
\;\; \psi^{ijI}:=B^{ijI}-F^{ijI}\approx 0,
\end{split}
\end{equation}
\noindent and the rank yields  fix the following values for the
Lagrange multipliers
\begin{equation}\begin{split}
\dot{\phi}^{0iI}&=\{\phi^{0iI},H_{P}\}\approx 0 \;\; \Rightarrow\;\;
\lambda_{i}^{I}=0,\\
\dot{\phi}^{ijI}&=\{\phi^{ijI},H_{P}\}\approx 0 \;\; \Rightarrow\;\;
\lambda_{0i}^{I}=2D_{j}B^{jiI}-2f^{IJK}A_{0}^{J}\Pi^{iK}.
\end{split}
\end{equation}
\noindent For this theory there are not third constraints, instead
we obtain the following  Lagrange multipliers.
\begin{equation}
\dot{\psi}^{lmP}=\{\psi^{lmP},H_{P}\}\approx 0 \;\; \Rightarrow\;\;
\alpha_{lm}^{P}=0,\;\;\;
\lambda_{lm}^{P}=D_{l}\Pi^{mP}-D_{m}\Pi^{lP}-f^{PKI}F_{lm}^{K}A_{0}^{I}
\end{equation}
\noindent In this manner, with all the constraints at hand , we need
identify those that are first and second class kind. For this
purpose, we can observe that the $14(N^{2}-1)\times 14(N^{2}-1)$
matrix whose entries are the Poisson�s brackets among the primary
and secondary constraints are given by
\begin{eqnarray}
\{ \phi^{0P}(x),\phi^{0I}(y) \}&=&0,   \qquad   \{ \phi^{0P}(x),\phi^{iI}(y) \} = 0,  \nonumber \\
\{ \phi^{0P}(x),\phi^{0iI}(y) \}&=&0,   \qquad   \{ \phi^{0P}(x),\phi^{ijI}(y) \} = 0,  \nonumber \\
\{ \phi^{lP}(x),\phi^{iI}(y) \}&=&0,   \qquad   \{ \phi^{lP}(x),\phi^{0iI}(y) \} = -\frac{1}{2}\delta^{i}_{l}\delta^{PI}\delta^{3}(x-y),  \nonumber \\
\{ \phi^{lP}(x),\phi^{ijI}(y) \}&=&0,   \qquad   \{ \phi^{lmP}(x),\phi^{ijI}(y) \} = 0,  \nonumber \\
\{ \phi^{0P}(x),\psi^{I}(y) \}&=&0,   \qquad   \{ \phi^{lP}(x),\psi^{I}(y) \} = f^{PIK}\Pi^{lK}\delta^{3}(x-y),  \nonumber \\
\{ \phi^{0lP}(x),\psi^{I}(y) \}&=&0,   \qquad   \{ \phi^{lmP}(x),\psi^{I}(y) \} = 0,  \nonumber \\
\{ \psi^{P}(x),\psi^{ijI}(y) \} &=& -f^{PIM}F_{ij}^{M},   \qquad   \{  \psi^{P}(x),\psi^{I}(y) \}=f^{PIK}\psi^{K}=0  \nonumber \\
\{ \phi^{0P}(x),\psi^{ijI}(y) \}&=&0,   \qquad   \{ \phi^{lP}(x),\psi^{ijI}(y) \} =\left(\delta^{l}_{j}\delta^{PI}\partial_{i}-\delta^{l}_{i}\delta^{PI}\partial_{j}+f^{IPK}(\delta^{l}_{i}A_{j}^{K}+\delta^{l}_{j}A_{i}^{K})\right)\delta^{3}(x-y),  \nonumber \\
\{ \psi^{lmP}(x),\phi^{ijI}(y) \}&=&(\delta^{i}_{l}\delta^{j}_{m}-\delta^{i}_{m}\delta^{j}_{l})\delta^{PI}\delta^{3}(x-y),   \qquad   \{ \phi^{0lP}(x),\psi^{ijI}(y) \} = 0,  \nonumber \\
\{ \psi^{lmP}(x),\phi^{0I}(y) \}&=&0, \nonumber 
\end{eqnarray}
\noindent has rank=$12(N^{2}-1)$ and $2(N^{2}-1)$ null-vectors. From
the null-vectors we identify the following $2(N^{2}-1)$ first class
constraints
\begin{equation} \label{fc bf}\begin{split}
\gamma^{0I}&=\Pi^{0I}\approx 0,\\
 \gamma^{I}&=D_{i}\Pi^{iI}+2f^{IJK}B_{0i}^{J}\Pi^{0iK}+f^{IJK}B_{ij}^{J}\Pi^{ijK}\approx 0.\\
\end{split}
\end{equation}
In particular, we would like to stress that the full structure of the Gauss constraint in (\ref{fc bf}) has not been reported in the literature. On the other hand, that Gauss constraint is  the full  generator of   $SU(N)$ transformations of the theory under study. \
\noindent Furthermore, the rank allow us to identify the following 
$12(N^{2}-1)$ second class constraints
\begin{equation}\label{sc bf}\begin{split}
\chi^{iI}&=\Pi^{iI}+B^{0iI}\approx 0,\\
\chi^{0iI}&=\Pi^{0iI}\approx 0,\\
\chi^{ijI}&=\Pi^{ijI}\approx 0,\\
\phi^{ijI}&=(B^{ijI}-F^{ijI}) \approx 0.
\end{split}
\end{equation}

\noindent Therefore, the counting of degrees of freedom is performed
as follows. There are  $20(N^{2}-1)$ phase space variables,
$2(N^{2}-1)$ independent first class constraints and $12(N^{2}-1)$
second class constraints, thus the theory given in (\ref{bf-ym}) has
$2(N^{2}-1)$ degrees of freedom just like [YM] theory.\\
\noindent The algebra among the constraints (\ref{fc bf}) and (\ref{sc bf}) is given by
\begin{eqnarray}
\{ \gamma^{0P}(x),\gamma^{0I}(y) \}&=&0,   \qquad   \{ \chi^{lP}(x),\gamma^{I}(y) \} = f^{PIK}\chi^{lK}=0,  \nonumber \\
\{ \gamma^{0P}(x),\chi^{iI}(y) \}&=&0,   \qquad   \{ \chi^{0lP}(x),\gamma^{I}(y) \} = f^{PIK}\chi^{0lP}=0,  \nonumber \\
\{ \gamma^{0P}(x),\chi^{0iI}(y) \}&=&0,   \qquad   \{ \chi^{lmP}(x),\gamma^{I}(y) \} = f^{PIK}\chi^{lmK}=0,  \nonumber \\
\{ \gamma^{0P}(x),\chi^{ijI}(y) \}&=&0,   \qquad   \{ \phi^{lmP}(x),\gamma^{I}(y) \} = f^{PIK}\phi^{lmK}=0,  \nonumber \\
\{ \gamma^{0P}(x),\phi^{ijI}(y) \}&=&0,   \qquad   \{ \gamma^{P}(x),\gamma^{I}(y) \} = f^{PIK}\gamma^{K}=0,  \nonumber \\
\{ \gamma^{0P}(x),\gamma^{I}(y) \}&=&0,   \qquad   \{ \chi^{lP}(x),\chi^{iI}(y) \} = 0,  \nonumber \\
\{ \chi^{lP}(x),\chi^{0iI}(y) \} &=& -\frac{1}{2}\delta^{i}_{l}\delta^{PI}\delta^{3}(x-y),   \qquad   \{  \chi^{lmP}(x),\chi^{ijI}(y) \}=0  \nonumber \\
\{ \chi^{lP}(x),\chi^{ijI}(y) \}&=&0,   \qquad   \{ \chi^{lP}(x),\phi^{ijI}(y) \} =\left(\delta^{l}_{j}\delta^{PI}\partial_{i}-\delta^{l}_{i}\delta^{PI}\partial_{j}+f^{PIK}(\delta^{l}_{j}A_{i}^{K}-\delta^{l}_{i}A_{j}^{K}\right)\delta^{3}(x-y),  \nonumber \\
\{ \chi^{lP}(x),\phi^{ijI}(y) \}&=&0,   \qquad   \{ \chi^{0lP}(x),\phi^{ijI}(y) \} = 0,  \nonumber \\
\{ \chi^{0lP}(x),\chi^{0iI}(y) \}&=&0,  \qquad   \{ \chi^{lmP}(x),\phi^{ijI}(y) \} = -\frac{1}{2}\left(\delta^{l}_{i}\delta^{m}_{j}-\delta^{l}_{j}\delta^{m}_{i}\right)\delta^{PI}\delta^{3}(x-y) ,  \nonumber \\
\{ \phi^{lmP}(x),\phi^{ijI}(y) \}&=&0. \nonumber 
\end{eqnarray}
\noindent where we can appreciate that the algebra  is closed.\\
\noindent The identification of the constraints will allow us to
identify the extended action. By using the first class constraints
(\ref{fc bf}), the second class constraints (\ref{sc bf}), and the Lagrange multipliers
we find that the extended action takes the form
\begin{align}
S_{E}& [A_{\mu}^{I},\Pi^{\mu I},B_{\mu\nu}^{I},\Pi^{\mu\nu
I},\lambda_{0}^{I},\lambda^{I},u_{i}^{I},u_{0i}^{I},u_{ij}^{I},v_{ij}^{I}]=\int
d^{4}x(\dot{A}_{\mu}^{I}\Pi^{\mu I}+\dot{B}_{\mu\nu}^{I}\Pi^{\mu\nu
I}-\frac{1}{2}\Pi^{iI}\Pi_{i}^{I} +\frac{1}{4}B_{ij}^{I}B^{ijI} \nonumber\\
+A_{0}^{I}&
D_{i}\Pi^{iI}-\frac{1}{2}B_{ij}^{I}F_{ij}^{I}-2D_{i}B^{ijI}\Pi^{0jI}+2f^{PKI}\Pi^{lK}A_{0}^{I}\Pi^{0lK}
-2D_{l}\Pi^{mP}\Pi^{lmP}+f^{PKI}F_{lm}^{K}A_{0}^{I}\Pi^{lmP}\nonumber\\
-\lambda_{0}^{I}&\gamma^{0I}-\lambda^{I}\gamma^{I}
-u_{i}^{I}\chi^{iI}-u_{0i}^{I}\chi^{0iI}-u_{ij}^{I}\chi^{ijI}-v_{ij}^{I}\phi^{ijI}).
\end{align}
\noindent From the extended action we can identify the extended
Hamiltonian given by
\begin{equation}
H_{E}=H+\lambda_{0}^{I}\gamma^{0I}+\lambda^{I}\gamma^{I},
\end{equation}
\noindent where $H$ is given by
\begin{align}
H& =\frac{1}{2}\Pi^{iI}\Pi_{i}^{I} +\frac{1}{4}B_{ij}^{I}B^{ijI}
+A_{0}^{I}
D_{i}\Pi^{iI}-\frac{1}{2}B_{ij}^{I}F_{ij}^{I}-2D_{i}B^{ijI}\Pi^{0jI}+2f^{PKI}\Pi^{lK}A_{0}^{I}\Pi^{0lK} \nonumber\\
-& 2D_{l}\Pi^{mP}\Pi^{lmP}+f^{PKI}F_{lm}^{K}A_{0}^{I}\Pi^{lmP}.
\end{align}
\noindent We will continue this section by computing the equations
of motion obtained from the extended action, which are expressed by
\begin{align}
\delta A_{0}^{P}&
:\dot{\Pi}^{0P}=D_{l}\Pi^{lP}+2f^{JKP}\left(\Pi^{lK}\Pi^{0J}+F_{lm}^{K}\Pi^{lmJ}\right),\nonumber\\
\delta\Pi^{0P}& :\dot{A}_{0}^{P}=\lambda_{0}^{P},\\
\delta A_{l}^{P}&
:\dot{\Pi}^{lP}=f^{IPK}A_{0}^{I}\Pi^{lK}+D_{i}B_{il}^{P}-2f^{KPJ}B^{ljJ}\Pi^{0jK}-2f^{IPK}\Pi^{jK}\Pi^{ljI}-2D_{i}\left(f^{KPI}A_{0}^{I}\Pi^{ilK}\right)
\nonumber\\ \quad & +f^{IPK}\Pi^{lK}\lambda^{I}+2D_{i}v^{ilP},\nonumber\\
\delta\Pi^{lP}&
:\dot{A}_{l}^{P}=\Pi^{lP}-D_{l}A_{0}^{P}-2f^{KPI}A_{0}^{I}\Pi^{0lK}+u^{lP}\nonumber\\
\delta B_{0l}^{P}&
:\dot{\Pi}^{0lP}=f^{PIK}\Pi^{0lK}\lambda^{I}-\frac{1}{2}u^{lP}
\nonumber \\
\delta B_{lm}^{P}&
:\dot{\Pi}^{lmP}=2D_{l}\Pi^{0mP}+f^{PIK}\Pi^{lmK}\lambda^{I}-\frac{1}{2}u_{lm}^{P}\nonumber\\
\delta \Pi^{0lP}&
:\dot{B}_{0l}^{P}=D_{i}B^{ilP}-f^{PKI}\Pi^{lK}A_{0}^{I}+\frac{1}{2}u_{0l}^{P}\nonumber\\
\delta\Pi^{lmP}&
:\dot{B}_{lm}^{P}=2D_{l}\Pi^{mP}-f^{PKI}F_{lm}^{K}A_{0}^{I}+f^{IJP}\lambda^{I}B_{lm}^{J}+u_{lm}^{P}\nonumber\\
\delta\lambda_{0}^{I}& :\gamma^{0I}=0\nonumber\\
\delta\lambda^{I}& :\gamma^{I}=0\nonumber\\
\delta u_{i}^{I}& :\chi_{i}^{I}=0\nonumber\\
\delta u_{0i}^{I}& :\chi_{0i}^{I}=0\nonumber\\
\delta u_{ij}^{I}& :\chi_{ij}^{I}=0\nonumber\\
\delta v_{i}^{I}& :\phi_{i}^{I}=0\nonumber
\end{align}
\noindent By following with our analysis, we need to know the gauge
transformations on the phase space of the theory under study. For
this step, we shall use  Castellani's formalism which allow us to
define the following gauge generator in terms of the first class
constraints (\ref{fc bf})
\begin{equation}
G=\int_{\Sigma}\left[D_{0}\epsilon_{0}^{I}\gamma^{0I}+\epsilon^{I}\gamma^{I}\right]d^{3}x,
\end{equation}
\noindent thus, we find that the gauge transformations on the phase
space are given by
\begin{align}
\delta_{0} A_{0}^{P}& =D_{0}\epsilon_{0}^{P},\nonumber\\
\delta_{0} A_{i}^{P}& =-D_{i}\epsilon^{P},\nonumber\\
\delta_{0} \Pi^{0P}& = -f^{PKI}\epsilon_{0}^{K}\Pi^{0I},\nonumber\\
\delta_{0} \Pi^{iP}& = f^{PIK}\Pi^{iK}\epsilon^{I},\nonumber\\
\delta_{0} B_{0i}^{P}& = f^{PIJ}\epsilon^{I}B_{0i}^{J},\nonumber\\
\delta_{0} \Pi^{0i}& = f^{PIK}\epsilon^{I}\Pi^{0iK},\nonumber\\
\delta_{0} B_{ij}^{P}& = f^{PIJ}\epsilon^{I}B_{ij}^{J},\nonumber\\
\delta_{0} \Pi^{ijP}& = f^{PIK}\epsilon^{I}\Pi^{ijK}.
\end{align}

\noindent We can observe that by redefining the gauge parameters
$\epsilon_{0}^{I}=\epsilon^{I}$, the gauge transformations take the
form

\begin{align}
A_{\mu}^{'I}& \rightarrow
A_{\mu}^{I}-D_{\mu}\epsilon^{I},\nonumber\\
B_{\mu\nu}^{'I}& \rightarrow B_{\mu\nu}^{I}-f^{I}{_{JK}}\epsilon^{J}B_{\mu\nu}^{K},
\end{align}
where the first one transformation corresponds to the usual gauge transformations for [YM] theory, and the later one by using the equations of motion gives us the transformation
of a valued compact Lie algebra curvature tensor field. In order to obtain the path integral quantization of the theory and its \textit{uv}  behaviour,  it is straightforward to perform  
Senjanovic's method  \cite{senjanovic} by  taking into account  the full constraint algebra obtained above to   define the corresponding non-abelian measure. After some integration 
over the second class constraints,  and by using the first class constraints  to identify an  appropriate gauge fixing \cite{ Weinberg},  one finally gets the usual quantum effective action 
of the [YM] theory.       
\section{ Martellini's model }
An interesting alternative model to express [YM] theory as a constrained $BF$-like theory has been reported  by  M. Martellini and M. Zeni \cite{Martellini 1}. Martellini's model is a deformation 
of a topological field theory, namely the pure $BF$ theory resulting in the first order formulation of [YM] theory. In this formulation, 
new non local observables can be introduced following the topological theory and giving an explicit realization of t'Hooft algebra, recovering at the end
the standard \textit{u-v} behaviour of the theory \cite{Martellini 2}. So, the aim of this section is to perform the canonical analysis for Martellini's model on the full phase space context,   which is absent in the literature,  then we compare the results obtained with those found in  former sections. \\
Let us  start with the action proposed by Martellini et al \cite{Martellini 1}
\begin{equation}
S[A,B]= \int \frac{i}{2} \varepsilon^{\mu \nu \alpha \beta} B_{I \mu
\nu}F{^{I}}_{\alpha \beta}+  g^2\int  B{^{I}}_{\mu \nu}B{_{I}}^{\mu
\nu}. \label{eq52}
\end{equation}
The firs term in the r.h.s. of (\ref{eq52}) is the usual $BF$ theory laking of local degrees of freedom, and has been analyzed within a smaller phase space context in  \cite{20}, and by using a pure Dirac's analysis in \cite{4a}. As we shall see below,  local degrees of freedom are restored by the $g^2 B{^{I}}_{\mu \nu}B{_{I}}^{\mu
\nu}$  term  of (\ref{eq52}),  allowing  an explicit
 breaking  of the topological sector as long as $\mathit{g}\neq 0$. Therefore, in Martellini's  formulation  [YM] theory is expressed  as a deformation of the
topological $BF$ field theory. \\
We are able to  observe that the actions (\ref{bf-ym})  and  (\ref{eq52})
differ in the first term. In (\ref{bf-ym}) neither is present  the
imaginary number that provides the euclidean feature nor the space time indices are contracted with the
epsilon tensor. However, because  the  physical relation among  [YM]  and the action
(\ref{eq52}), in this section we are interested in develop   a complete  Hamiltonian framework of  the action (\ref{eq52}) because is   absent in the literature. \\
By  performing  the $3+1$ decomposition of the action (\ref{eq52}) we obtain
\begin{equation}
S[A,B]=\int\int_{\Sigma}dtd^{3}xg^{2}\left(2B_{0i}^{I}B^{0iI}+B_{ij}^{I}B^{ijI}\right)+i\eta^{ijk}\left(B_{0i}^{I}F_{jk}^{I}+B_{ij}^{I}F_{0k}^{I}\right),
\end{equation}
hence, by following the procedure developed in above section, we find  the following results;  there are the following  $2(N^{2}-1)$ first class
constraints
\begin{equation} \label{Mfcc}\begin{split}
\gamma^{0I}&=\Pi^{0I}\approx 0,\\
 \gamma^{I}&=D_{i}\Pi^{iI}+2f^{IJK}B_{0i}^{J}\Pi^{0iK}+f^{IJK}B_{ij}^{J}\Pi^{ijK}\approx 0,
\end{split}
\end{equation}
and $12(N^{2}-1)$ second class constraints
\begin{equation}\label{Mscc}\begin{split}
\phi^{iI}&=\Pi^{iI}-i\eta^{ijk}B_{jk}^{I}\approx 0,\\
\phi^{0iI}&=\Pi^{0iI}\approx 0,\\
\phi^{ijI}&=\Pi^{ijI}\approx 0,\\
\psi^{0iI}&=2g^{2}B^{0iI}+\frac{i}{2}\eta^{ijk}F_{jk}^{I}\approx 0.
\end{split}
\end{equation}
\noindent Therefore, the counting of degrees of freedom is performed
as follows. There are  $20(N^{2}-1)$ phase space variables,
$2(N^{2}-1)$ independent first class constraints and $12(N^{2}-1)$
second class constraints, thus the theory given in (\ref{eq52}) has
$2(N^{2}-1)$ degrees of freedom.\\
Now, we observe that the algebra of the constraints is given by
\begin{eqnarray}
\{ \gamma^{0P}(x),\gamma^{0I}(y) \}&=&0,   \qquad   \{ \phi^{lP}(x),\gamma^{I}(y) \} = f^{PIK}\phi^{lK}=0,  \nonumber \\
\{ \gamma^{0P}(x),\phi^{iI}(y) \}&=&0,   \qquad   \{ \phi^{0lP}(x),\gamma^{I}(y) \} = f^{PIK}\phi^{0lP}=0,  \nonumber \\
\{ \gamma^{0P}(x),\phi^{0iI}(y) \}&=&0,   \qquad   \{ \phi^{lmP}(x),\gamma^{I}(y) \} = f^{PIK}\phi^{lmK}=0,  \nonumber \\
\{ \gamma^{0P}(x),\phi^{ijI}(y) \}&=&0,   \qquad   \{ \psi^{0lP}(x),\gamma^{I}(y) \} = f^{PIK}\psi^{0lK}=0,  \nonumber \\
\{ \gamma^{0P}(x),\psi^{0iI}(y) \}&=&0,   \qquad   \{ \gamma^{P}(x),\gamma^{I}(y) \} = f^{PIK}\gamma^{K}=0,  \nonumber \\
\{ \gamma^{0P}(x),\gamma^{I}(y) \}&=&0,   \qquad   \{ \phi^{lP}(x),\phi^{iI}(y) \} = 0,  \nonumber \\
\{ \phi^{lP}(x),\phi^{0iI}(y) \} &=& 0,   \qquad   \{  \phi^{lmP}(x),\phi^{ijI}(y) \}=0  \nonumber \\
\{ \phi^{lP}(x),\phi^{ijI}(y) \}&=&-i\eta^{lij}\delta^{PI}\delta^{3}(x-y),   \qquad   \{ \phi^{0lP}(x),\psi^{0iI}(y) \} =-g^{2}\delta^{l}_{i}{\delta}^{PI}\delta^{3}(x-y),  \nonumber \\
\{ \phi^{lP}(x),\psi^{0iI}(y) \}&=& i\eta^{ijl}(\delta^{PI}\partial_{j}+f^{PIK}A_{j}^{K})\delta^{3}(x-y),   \qquad   \{ \phi^{0lP}(x),\phi^{ijI}(y) \} = 0,  \nonumber \\
\{ \phi^{0lP}(x),\phi^{0iI}(y) \}&=&0,  \qquad   \{ \psi^{lmP}(x),\psi^{ijI}(y) \} = 0 ,  \nonumber 
\end{eqnarray}
\noindent where we can appreciate that the constraints form a set of
first and second class constraints as is expected.  \\
On the other hand, the identification of the constraints will allow us to
identify the extended action. By using those results,   we find the extended action given by 
\begin{align}
S_{E}& [A_{mu}^{I},\Pi^{\mu I},B_{\mu\nu}^{I},\Pi^{\mu\nu
I},\lambda_{0}^{I},\lambda^{I},u_{i}^{I},u_{0i}^{I},u_{ij}^{I},v_{0i}^{I}]=\int
d^{4}x(\dot{A}_{\mu}^{I}\Pi^{\mu I}+\dot{B}_{\mu\nu}^{I}\Pi^{\mu\nu
I}-\frac{1}{2}\Pi^{iI}\Pi_{i}^{I} + g^{2}2B_{0i}^{I}B^{0iI} \nonumber\\
+A_{0}^{I}& D_{i}\Pi^{iI}+i\eta^{ijk}B_{0i}^{I}F_{jk}^{I}
-\frac{1}{2g^{2}}\eta^{ijk}f^{PIJ}A_{0}^{I}F_{jk}^{J}\Pi^{0iP}+\frac{i}{2}\eta^{ijk}D_{j}\Pi^{kP}\Pi^{0iP}-2D_{i}B_{0j}^{I}\Pi^{ijI} \nonumber\\
-\frac{i}{2}&\eta_{ijk}f^{PIK}A_{0}^{I}\Pi^{kK}\Pi^{ijP}-\lambda_{0}^{I}\gamma^{0I}-\lambda^{I}\gamma^{I}
-u_{i}^{I}
\phi^{iI}-u_{0i}^{I}\phi^{0iI}-u_{ij}^{I}\phi^{ijI}-v_{0i}^{I}\psi^{0iI}).
\end{align}
\noindent From the extended action we can identify the extended
Hamiltonian  given by
\begin{equation}
H_{E}=H+\lambda_{0}^{I}\gamma^{0I}+\lambda^{I}\gamma^{I},
\end{equation}
\noindent where $H$ has the following form
\begin{align}
H& =\frac{1}{2}\Pi^{iI}\Pi_{i}^{I}
-2g^{2}B_{0i}^{I}B^{0iI}-A_{0}^{I}
D_{i}\Pi^{iI}-i\eta^{ijk}B_{0i}^{I}F_{jk}^{I}
+\frac{1}{2g^{2}}\eta^{ijk}f^{PIJ}A_{0}^{I}F_{jk}^{J}\Pi^{0iP}-\frac{i}{2}\eta^{ijk}D_{j}\Pi^{kP}\Pi^{0iP}\nonumber\\
&+2D_{i}B_{0j}^{I}\Pi^{ijI}
+\frac{i}{2}\eta_{ijk}f^{PIK}A_{0}^{I}\Pi^{kK}\Pi^{ijP}.
\end{align}
Hence, the
following question rise;  Are there differences among the action (\ref{bf-ym1}) and Martellini's propose?.  The difference   lies in the constraint algebra, in fact, 
 we  observe the algebra among the second class constraints for action (\ref{bf-ym1}) and  Martellini's is different. Furthermore, in   Martellini's  theory the algebra among the constraints  is defined over the complex numbers, consequence of a Wick rotation, and the definition of the momenta gives dual expressions of the constraints defined for the action (\ref{bf-ym1}). In
particular note that in Martellini's model,   $B$ is proportional to the field strength and satisfies the Bianchi identities on-shell. This is no longer true off-shell and this fact
has been related to the presence of monopole charges in the vacuum \cite{Martellini 2} which should enter in the non perturbative sector of the theory. Moreover
the action (\ref{eq52}) has been  used to define new non local observables related to the phase space structure of the theory \cite{Cattaneo 2}. On the other side, it is mandatory to investigate 
the quantum behavior of the action (\ref{bf-ym1}) at perturbative level for finding new local observables,   and  thus, compare with Martellini's model possibles advantages of the action  (\ref{bf-ym1}); we remark  that  the action (\ref{bf-ym1}) and  Martellini's model   have  different  algebra among  the second class constraints,  and this fact will be important in the quantum treatment for instance, in the construction of Dirac's brakets. In this respect, the present letter   has  the necessary tools for studying  these subjects in forthcoming  works. 
\section{ Conclusions and prospects}
In this paper, we have developed  a consistent application of a pure Dirac's method for constrained systems. By working with the original phase space we performed a complete 
Hamiltonian dynamics for two $BF$-like theories. The first one,  was related with [YM] theory,   and the second  action was associated with  Martellini's model, which 
has been used in recently works for studying  the non perturbative character of the QCD confinement. From the present analysis,  we calculated for the  theories under study, the extended action, 
the extended Hamiltonian and the full constraints program, which is considerably enlarged in comparison with the analysis performed on the reduce phase space.  The correct identification of the constraints  as first and second class, enabled us to carry out the counting of degrees of freedom, concluding that classically, the theories under study have the same number of degrees  of freedom of  [YM] theory. The full phase space framework,  allowed us  observe  that the physical degrees of freedom emerge from the coupling of  topological theories. The topological  invariance  is broken because there are not in the full action reducibility conditions among the constraints, which endow the theory with local dynamics. The nature of such conditions are closely related to the full phase space, and cannot be obtained from the reduce one.    
With regard to the quantum aspect, the application of the pure Dirac's procedure provides the full structure of the constraints, and this fact give us  a complete gauge information of the theory. It is worth mentioning that once the full set of constraints is calculated, our  procedure could shed light on the search of observables in the context of covariant field theories  specifically in the case of strong-Dirac observables, which must be defined in the complete phase space. Finally, we observed that  the  action (\ref{bf-ym1}) and Martellini's model yield [YM] equations of motion, however, the algebra of their   constraints is different, thus, we expect different quantum scenarios for these theories, all those ideas are in progress and will be reported in forthcoming works.    \\

\noindent \textbf{Acknowledgements}\\[1ex]
This work was supported by CONACyT  M\'exico under grant  157641.
 

\begin{thebibliography}{100}
\setlength{\itemsep}{-.50em}
\bibitem{carlo}  C. Rovelli, Quantum Gravity (Cambridge University
Press, Cambridge, England, 2004)
\bibitem{tomas}  T.Thiemann, Modern Canonical Quantum General Relativity (Cambridge, UK: Cambridge Univ. Pr. 2007  )
\bibitem{3a} M. Montesinos and  A. Perez,  Phys.Rev.D77:104020,2008.
\bibitem{5a} G.T Horowitz, Commun. Math. Phys. 125, 417,  (1989).
\bibitem{6a} G.T Horowitz,  M. Srednicki,  Commun. Math. Phys. 130, 83,  (1990).
\bibitem{baez} J. Baez,  Lect.Notes Phys. 543 (2000) 25-94.
\bibitem{8aa} J. F. Plebanski, J. Math. Phys. 18, 2511 (1977).
\bibitem{7a} Derek K. Wise, {\it Macdowell Mansouri Gravity and Cartan Geometry, Available from: hep-th/0501191}
\bibitem{8a} A. Accardi, A. Belli, M. Martellini and M. Zeni, hepth/9703152.
\bibitem{Martellini 1} F. Fucito, M. Martellini and M. Zeni, hep-th/9605018.
\bibitem{Cattaneo} A.S. Cattaneo, P. Cotta Ramusino, A. Gamba and M. Martellini. Phys. Lett. B355 (1995) 245.
\bibitem{Martellini 2} F. Fucito, M. Martellini and M. Zeni, hep-th/9607044.

\bibitem{Bertrand} B. Bertand, J. Govaerts, hep-th/0704.1512v1.
\bibitem{Bertrand1} B. Bertrand, J. Govaerts, hep-th/0705.3452v1.
\bibitem{Freidel} L. Freidel, K. Krasnov, arXiv:0708.1595 [gr-qc].
\bibitem{locality} G. Amelino-Camelia, L. Freidel, J. Kowalski-Glikman, L. Smolin, hep-th:1101.093.
\bibitem{Pullin}C. Di Bartolo, R. Gambini, J. Pullin, Class. Quan. Grav. 19, 5475 (2002).
\bibitem{Lisi} A. G. Lisi, {\it An Exceptionally Simple Theory of Everything}, arXiv:0711.0770.
\bibitem{Smolin} L. Smolin, {\it The Plebanski action extended to a unification of gravity and [YM] theory}, arXiv:0712.0977v2.
\bibitem{4bb} A. Escalante and Leopoldo Carbajal, Annals of Physics 326, 323-339, (2011). 
\bibitem{4aa} A. Escalante, Int. J. Theo. Phys, Vol 48, No. 9, 2473-2729. (2009).
\bibitem{4a} A. Escalante and Ira{\'i}s Rubalcava, {\it A Pure Dirac's method for 4-dimensional BF theories},  to be published in Int. J. Geom. Methods Mod. Phys  (2012). 
\bibitem{Peldan} P. Peldan, {\it Actions for Gravity, with Generalizations: A review }, arXiv:gr-qc/930511v1.
\bibitem{ww} E.Buffenoir, M.Henneaux, K.Noui, Ph.Roche, Class.Quant.Grav. 21 (2004) 5203-5220.
\bibitem{Witten}  A. M. Frolov, N. Kiriushcheva and S. V. Kuzmin, Gravitation and Cosmology, 16: 181-194, (2010);  E. Witten, Nucl. Phys. B 311,  46-78, (1988) .
\bibitem{Carlip} S. Carlip, Phys. Rev. D 42 (1990) 2647-2654.
\bibitem{21a} A. Escalante, Phys.Lett.B,  676:105-11, (2009).
\bibitem{22a} M. Montesinos, Class.Quant.Grav.23:2267-2278, (2006). 
\bibitem{senjanovic} Senjanovic P. Ann. Phys. (N.Y.) 100, 227 (1976)
\bibitem{Weinberg} S. Weinberg, The Quantum Theory of Fields Vol. II, Cambridge University Press, 1996. 
\bibitem{20} M. Mondragon and M. Montesinos, J. Math. Phys.
47, 022301 (2006).
\bibitem{Cattaneo 2} A. S. Cattaneo, P. Cotta-Ramusino, J. Frohlich and M. Martellini, J. Math. Phys. 36 (1995) 6137.
\bibitem{Martellini 3} M. Martellini and M. Zeni, hep-th/9610090.
\end{thebibliography}
\end{document}